\documentclass[a4paper,twocolumn]{esapub}
\usepackage[T1]{fontenc}
\usepackage[latin1]{inputenc}
\usepackage{array}
\usepackage{amsfonts}
\usepackage{amsmath}
\usepackage{amssymb}
\usepackage{amsthm}
\usepackage{color}
\usepackage{times}
\usepackage{natbib}
\usepackage{graphicx}

\providecommand{\tabularnewline}{\\}

\title{Systematic Characterization of Low Frequency Electric and Magnetic Field Data Applicable to Solar Orbiter}
\author{Jan~E.~S.~Bergman}
\affil{Swedish Institute of Space Physics, P.O. Box 531, SE-751 21  Uppsala, Sweden, E-mail: jb@irfu.se}
\author{Tobia~D.~Carozzi}
\affil{Space Science Centre, Sussex University, Falmer, E. Sussex, BN1 9QT, UK,
E-mail: T.Carozzi@sussex.ac.uk}

\begin{document}

\maketitle

\newcommand{\dual}{\vphantom{F}^{\ast}\!}
\newcommand{\blank}[1]{\phantom{#1}}
\newcommand{\CompConj}[1]{\bar{#1}}
\newcommand{\CompConjW}[1]{\overline{#1}}
\newcommand{\icu}{\mathbf{i}}
\newcommand{\selfDual}{\vphantom{F}^{{\scriptscriptstyle {+}}}\!}
\newcommand{\selfDualAnti}{\vphantom{F}^{{\scriptscriptstyle {-}}}\!}
\newcommand{\tensor}[1]{\bf{\sf{{#1}}}}
\renewcommand{\vec}[1]{\mathbf{{#1}}}


\begin{abstract}
We present a systematic and physically motivated characterization of incoherent
or coherent electric and magnetic fields, as measured for instance by the low
frequency receiver on-board the Solar  Orbiter spacecraft. The characterization
utilizes the $36$ auto/cross correlations of the $3+3$ complex Cartesian components
of the electric and magnetic fields; hence, they are second order in the field
strengths and so have physical dimension energy density.
Although such $6\times6$ correlation matrices have been successfully
employed on previous space missions,
they are not physical quantities; because they are not manifestly space-time tensors.
In this paper we propose a systematic representation of the $36$ degrees-of-freedom
of partially coherent electromagnetic fields as a set of manifestly covariant
space-time tensors,
which we call the Canonical Electromagnetic Observables (CEO).
As an example, we apply this formalism to analyze real data from
a chorus emission in the mid-latitude
magnetosphere, as registered by the STAFF-SA instrument on board the Cluster-II spacecraft.
We find that the CEO analysis increases the amount of information
that can be extracted from the STAFF-SA dataset; for instance, the reactive energy flux
density, which is 
one of the CEO parameters, identifies the source region of electromagnetic emissions
more directly than the active energy (Poynting) flux density alone.
\end{abstract}

\section{Background}
When analyzing time varying electric and magnetic vector field data from spacecraft, 
it is common 
to construct a $6\times6$ matrix from the complex sixtor 
$(\vec{E},\vec{B})$. In a Cartesian coordinate system, this 
matrix 
can be written
\begin{align}
\left(
\begin{array}{cccccc}
|E_x|^2 & E_x E_y^{\star} & E_x E_z^{\star} & E_x B_x^{\star} & E_x B_y^{\star} & E_x B_z^{\star} \\
E_y E_x^{\star} & |E_y|^2 & E_y E_z^{\star} & E_y B_x^{\star} & E_y B_y^{\star} & E_y B_z^{\star} \\
E_z E_x^{\star} & E_z E_y^{\star} & |E_z|^2 & E_z B_x^{\star} & E_z B_y^{\star} & E_z B_z^{\star} \\
B_x  E_x^{\star} & B_x E_y^{\star} & B_x E_z^{\star} & |B_x|^2 & B_x B_y^{\star} & B_x B_z^{\star} \\
B_y E_x^{\star} & B_y E_y^{\star} & B_y E_z^{\star} & B_y B_x^{\star} & |B_y|^2 & B_y B_z^{\star} \\
B_z E_x^{\star} & B_z E_y^{\star} & B_z E_z^{\star} & B_z B_x^{\star} & B_z B_y^{\star} & |B_z|^2
\nonumber
\end{array}
\right)
\end{align}
This electromagnetic (EM) sixtor matrix has in various guises, such as 
Wave-distribution functions (WDF) \cite{Storey:1974} and so on, been useful 
in the analysis of EM vector field data from spacecraft, such as on the
Cluster and Polar missions.
Although this matrix-form description of the second order properties
of the EM fields can in some instances be convenient and intuitive, it unfortunately
has no real motivation in physics, since physical quantities such as EM fields
are ultimately not sixtors.

The EM sixtor matrix lists all possible auto and cross correlations 
of $\vec{E}(\vec{x},t)$ and $\vec{B}(\vec{x},t)$ and, hence, 
contains the complete information
of the second order properties of EM waves.
It is clear that one obtains the EM energy 
density\footnote{Throughout the paper, when second order field quantities 
are discussed we have chosen a normalization such that 
the speed of light is set to unity.} $|\vec{E}|^2+|\vec{B}|^2$ by taking the trace of the EM sixtor matrix 
but it is not clear what other EM quantities can be extracted 
and how this extraction should be performed in general. Some EM quantities 
are obvious and can
be picked out by hand, such as the Lagrangian density 
$|\vec{E}|^2-|\vec{B}|^2$, the Poynting flux density 
$\Re\{\vec{E}\times\CompConj{\vec{B}}\}$, etc; 
$\CompConj{\vec{B}}$ denotes the
complex conjugate of $\vec{B}$, etc. Other EM quantities are not 
so easily identified. Furthermore, as it stands here, the EM sixtor
matrix is a mixture of both scalar and pseudo scalar components. This is due to
$\vec{E}$ being a proper (polar) vector and $\vec{B}$ being a pseudo (axial)
vector. This could be remedied by for instance using the sixtor 
$(\vec{E},\pm\icu\vec{B})$ but then again, it is unclear what 
sign to use for the imaginary unit, $\icu$. 
Regardless of how the sign of $\icu$ is
chosen one would need to redefine
the EM quantities. The trace of the modified proper EM sixtor matrix would 
in this case correspond to the
Lagrangian density rather than the EM energy density, which in turn
would have to
be redefined.
Other EM quantities would 
also need to be redefined in non-standard ways. 

Perhaps even more important for space borne observations: the EM
sixtor matrix is not covariant according to the requirements
of special relativity. Spacecraft are constantly moving and often spinning
observation platforms.
EM wave measurements become Doppler shifted and data must 
often be ``despun''.    
For the $3\times 3$ sub-matrices, 
$\vec{E}\otimes\CompConj{\vec{E}}$,
$\vec{E}\otimes\CompConj{\vec{B}}$,
$\vec{B}\otimes\CompConj{\vec{E}}$, and
$\vec{B}\otimes\CompConj{\vec{B}}$, where $\otimes$ denote the direct product,
despinning is straight forward by
applying rotation matrices $\tensor{R}$ from left and 
$\tensor{R}^{\rm{T}}$ from right, 
\emph{e.q.} 
$\vec{E}'\otimes\CompConj{\vec{E}}'
=\tensor{R}\vec{E}\otimes\CompConj{\vec{E}}\tensor{R}^{\rm{T}}$.
To rotate the full EM sixtor matrix, similar operations must be
performed four times. This is awkward and the resulting $6\times 6$ matrix
is still not covariant. 
For EM wave measurements in space plasma the last remark
can be crucial. 

A Lorentz boost is the
translation from one Lorentz frame to another one moving at 
velocity $\vec{v}$. A Lorentz boost does not
necessarily imply
relativistic speeds, which is a common misconception; and therefore it do not by itself
preclude what is typically associated with relativistic effects.
It is simply a quite general recipe
to make two different observers agree on a physical observation.  
The Lorentz boost of the EM
field vectors can be written 
$\vec{E}'=\gamma(\vec{E}+\vec{v}\times\vec{B})$ and
$\vec{B}'=\gamma(\vec{B}-\vec{v}\times\vec{E}/c^2)$,
where $c$ is the speed of light and $\gamma=1/\sqrt{1-v^2/c^2}$.
As a matter of fact, the Lorentz boost is the essence of the well-known 
frozen-in field line
theorem\footnote{If $\vec{E}+\vec{v}\times\vec{B}=0$ in a plasma, 
the magnetic field lines change as though they are convected with
velocity $\vec{v}$, \emph{i.e.}, they are frozen to the plasma flow. This
is the frozen-in field line theorem of ideal MHD.} 
from magnetohydrodynamics (MHD); a theory which is
commonly used to model the solar wind plasma. In a plasma,
relativity comes 
into play at very a fundamental level since the electromagnetic 
(Lorentz) force
dominates the vast majority of all plasma interactions.

Another example illustrates the problem to separate time (frequency) 
and 
space (wave vector) in EM wave observations 
on board a spacecraft. Assume that we observe a wave mode which 
is described
by an angular frequency $\omega$ and wave vector $\vec{k}$.
We can write this
as a 4-vector $(\omega,c\vec{k})$. Let's make a Lorentz boost in the
$\vec{v}$ direction:
\begin{eqnarray}
\omega'&=&\gamma(\omega-\vec{k}\cdot\vec{v})
\label{eq:omega_boost}\\
c\vec{k}'&=&c\vec{k} +\left[\frac{\gamma-1}{v^2}(c\vec{k}\cdot\vec{v})
-\gamma\omega\right]\frac{\vec{v}}{c}
\label{eq:ck_boost}
\end{eqnarray}
What happens now for a stationary (DC) field structure moving with the solar wind plasma? 
We then have $\omega=0$ and
$|c\vec{k}|\neq 0$. For a satellite moving with velocity $\vec{v}$ relative to
the DC field structure, it is justified to 
set $\gamma \approx 1$ (the solar wind speed seldom reaches more than
$900$ km/s and  using this value we obtain $\gamma\approx1.0000045\gtrsim 1$); Eqs.
(\ref{eq:omega_boost}) and
(\ref{eq:ck_boost}) are then reduced to
\begin{eqnarray}
\omega'&\approx&-\vec{k}\cdot\vec{v}
\label{eq:omega_boost_DC}\\
c\vec{k}'&\approx&c\vec{k}
\label{eq:ck_boost_DC}
\end{eqnarray}
We can see that the DC field structure is not Lorentz contracted appreciably at 
this low velocity,
$\vec{k}'\approx\vec{k}$. However, there is a dramatic change in the observed
frequency, which for a head-on encounter with the structure is registered as 
$\omega'\approx k v$ rather than zero. 
The observed frequency is proportional to the dimension of the structure, 
which we take to be in the order of one wavelength, $\lambda=2\pi/k$. 
Taking $v=900$ km/s a 900 km DC field structure would now register
as 1 Hz, a 90 km structure as 10 Hz, and a 9 km structure as 100 Hz, etc. 

These
simple examples clearly show that 
a space-time (covariant) description is necessary even if $\gamma\gtrsim 1$. 
The frequency (time) and the associated wave vector (space)
can not be treated separately but must be considered together, as a space-time
4-tensor.
 
The Maxwell equations are 
inherently relativistic and
can easily be put into a covariant form using 4-tensors. 
From a theoretical point of view,  
this fact alone provides a very good argument why 
one should try to
express also the second order properties of the EM fields
using a covariant formalism.
This was recently carried out by the authors and published in a recent paper 
\cite{Carozzi&Bergman:JMP:2006}. In this paper we
introduced a complete set of space-time tensors, 
which can fully describe the second order
properties of EM waves. We call this set of tensors the 
Canonical Electromagnetic Observables
(CEO); in analogy with Wolf's analysis of the Stokes parameters \cite{Wolf54}.
We suggest that the CEO could be used as an alternative to
the EM sixtor matrix. Not only are the CEO covariant, but they are all real valued
and provide a useful decomposition of the sixtor matrix into convenient 
physical quantities,
especially in the three-dimensional (3D), 
so-called scalar-vector-tensor (SVT) classification; 
see section \ref{SVTform}. 
The CEO have all dimension energy density but
have various physical interpretations as will be discussed in what follows.

\section{Canonical Electromagnetic Observables}
The CEO set was derived from the 
complex Maxwell field strength
$F^{\alpha\beta}$.
Other possibilities, such as using the 4-potential $A^{\mu}$
or using a spinor formalism \cite{Barut80}, were considered but
discarded due to their lack
of physical content. The 4-potential is not directly measurable and it is
furthermore gauge dependent. Spinor formalism has been proved possible
to use \cite{Sundkvist:2006} but we believe the space-time tensor
formalism to be more intuitive and convenient to use.

In the quantum theory of light, observables of
an EM field are ultimately constructed from a complex field strength; see \cite{Wolf54}. 
The simplest of these observables are sesquilinar-quadratic 
(Hermitian quadratic) 
in $F^{\alpha\beta}$, \emph{i.e.}, they are
functions of the components of $F^{\alpha\beta} \CompConj{F}^{\gamma\delta}$,
which is a 4-tensor of rank four. 
We showed that it was possible to
decompose 
$F^{\alpha\beta} \CompConj{F}^{\gamma\delta}$ 
into a unique set of tensors, the CEO, which
are real irreducible under the full Lorentz group.
We shall not repeat the derivation here but will instead discuss
the space-time (4-tensor) and three-dimensional (3-tensor) representations of the CEO. 

\subsection{Fundamental space-time representation}
\label{sec:4D}
In terms of the Maxwell field strength $F^{\alpha\beta}$, 
the CEO are organized
in the six real irreducible 4-tensors 
$C_{+},C_{-},Q^{\alpha\beta},T^{\alpha\beta},U^{\alpha\beta}$, and
$W^{\alpha\beta\gamma\delta}$. This is the fundamental space-time
representation of the CEO; 
their properties are listed in 
Table \ref{tab:4-tensor}.

\begin{table}[t]
\begin{tabular*}{\columnwidth}{@{\extracolsep{\fill}}|l|c|c|c|}
\hline
CEO & Rank &Proper $+$ & Number of\\
4-tensor & (Symmetry) &Pseudo $-$ & observables  \\
\hline
$C_+$ & 0 & $+$ & 1 \\
\hline
$C_-$ & 0 & $-$ & 1 \\
\hline
$T$ & 2(S) & $+$ & 9 \\
\hline
$U$ & 2(S) & $-$ & 9 \\
\hline
$Q$ & 2(A) & $-$ & 6 \\
\hline
$W$ & 4(M) & $+$ & 10 \\
\hline
\end{tabular*}
\label{tab:4-tensor}
\caption{CEO in space-time classification, \emph{i.e.}, 4-tensor notation:
$1+1+9+9+6+10=36$ observables.}
\end{table}

The CEO 4-tensors are defined as follows:
the two scalars are the vacuum proper- and pseudo-Lagrangians,
\begin{align}
C_{+}:= & \left(\CompConj{F}_{\alpha\beta}F^{\alpha\beta}
-\dual\CompConj{F}_{\alpha\beta}\dual F^{\alpha\beta}\right)/2\rm{,}
\label{eq:C+}\\
C_{-}:= & \left(\CompConj{F}_{\alpha\beta}\dual F^{\alpha\beta}
+\dual\CompConj{F}_{\alpha\beta}F^{\alpha\beta}\right)/2\rm{,}
\label{eq:C-}
\end{align}
respectively,
where we have used the dual of $F^{\alpha\beta}$ defined as\begin{equation}
\dual F^{\alpha\beta}:=\frac{1}{2}\epsilon^{\alpha\beta\gamma\delta}
F_{\gamma\delta}=\frac{1}{2}
\epsilon_{\blank{{\alpha}}\blank{{\beta}}\gamma\delta}^{\alpha\beta}
F^{\gamma\delta}.\label{eq:bivecDual}\end{equation}

The three second rank tensors consist of 
the two symmetric tensors
\begin{align}
T^{\alpha\beta}:= & \left(\CompConj{F}_{\blank{\alpha}\mu}^{\alpha}F^{\mu\beta}+\dual\CompConj{F}_{\blank{\alpha}\mu}^{\alpha}\dual 
F^{\mu\beta}\right)/2\rm{,}\label{eq:T-4}\\
U^{\alpha\beta}:= & \icu\left(\CompConj{F}_{\blank{\alpha}\mu}^{\alpha}\dual F^{\mu\beta}-\dual\CompConj{F}_{\blank{\alpha}\mu}^{\alpha}F^{\mu\beta}
\right)/2\rm{,}\label{eq:U-4}
\end{align}
and the antisymmetric tensor
\begin{align}
Q^{\alpha\beta}:= & \icu\left(\CompConj{F}_{\blank{\alpha}\mu}^{\alpha}F^{\mu\beta}-\dual\CompConj{F}_{\blank{\alpha}\mu}^{\alpha}\dual F^{\mu\beta}-2C_{+}\eta^{\alpha\beta}\right)/2\rm{.}\label{eq:Q-4}
\end{align}

The symmetric second rank tensor $T^{\alpha\beta}$ is the well-known EM
energy-stress tensor,
which contains the total energy, flux (Poynting vector),
and stress (Maxwell stress tensor)
densities.
The other two second rank tensors, $U^{\alpha\beta}$ and $Q^{\alpha\beta}$, respectively,
are less well-known. The symmetric  $U^{\alpha\beta}$ tensor is similar to 
$T^{\alpha\beta}$ in that it contains active energy densities but in $U^{\alpha\beta}$  
these densities are weighted and depend on the the handedness 
(spin, helicity, polarization, chirality)
of the EM field. 
Therefore, 
we have chosen to call them ``handed''  
energy densities. The anti-symmetric tensor $Q^{\alpha\beta}$ on the other hand is
very different in that it only contains reactive energy densities, which are
both total (imaginary part of the complex Poynting vector) and handed.

The fourth rank tensor is
\begin{eqnarray}
W^{\alpha\beta\gamma\delta}&:=& \left(\CompConj{F}^{\alpha\beta}F^{\gamma\delta}-\dual\CompConj{F}^{\alpha\beta}\dual F^{\gamma\delta}\right)/2
-2\icu Q^{[\alpha[\delta}\eta^{\gamma]\beta]}\nonumber\\
&-&\frac{2}{3}C_{+}\eta^{\alpha[\delta}\eta^{\gamma]\beta}
-\frac{1}{3}C_{-}\epsilon^{\alpha\beta\gamma\delta}
\label{eq:W}
\end{eqnarray}
where the square brackets denotes antisymmetrization over the enclosed
indices, \emph{e.g.}, $T^{\alpha[\delta}g^{\gamma]\beta}=\frac{1}{2}\left(T^{\alpha\delta}g^{\gamma\beta}-T^{\alpha\gamma}g^{\delta\beta}\right)$,
and nested brackets are not operated on by enclosing brackets, \emph{e.g.},
$T^{[\alpha[\delta}g^{\gamma]\beta]}=\frac{1}{4}\left(T^{\alpha\delta}g^{\gamma\beta}-T^{\alpha\gamma}g^{\delta\beta}-T^{\beta\delta}g^{\gamma\alpha}+T^{\beta\gamma}g^{\delta\alpha}\right)$.
It fulfills the symmetries $W^{\alpha\beta\gamma\delta}
=W^{\beta\alpha\gamma\delta}
=W^{\alpha\beta\delta\gamma}
=W^{\gamma\delta\alpha\beta}$ and $W^{\alpha[\beta\gamma\delta]}=0$.

This real irreducible rank four tensor, Eq. (\ref{eq:W}), was discovered by 
us\footnote{To the best of our knowledge, the $W^{\alpha\beta\gamma\delta}$ tensor
has never before been published in the literature.} and published in
\cite{Carozzi&Bergman:JMP:2006}, and is still under investigation; 
it is an extremely interesting geometrical object, having a structure identical
to the Weyl tensor in general relativity; see \cite{Weinberg1972}.
We have found
that it contains  
a four-dimensional generalization
of the Stokes parameters, 
as will be demonstrated in section \ref{sec:2D}
for the two-dimensional (2D)  case.
It contains both reactive total and reactive handed
energy densities.

\subsection{Three-dimensional representation}
\label{SVTform}
\begin{table}[t]
\begin{tabular*}{\columnwidth}{@{\extracolsep{\fill}}|l|c|c|c|}
\hline
& Scalars & Vectors & Tensors\\
\hline
(active) total & $u$  & $\vec{N}$ & ${\tensor{M}}$ \\
\hline
(active) handed & $h$  & $\vec{S}$ & ${\tensor{C}}$ \\
\hline
reactive total& $l$  & $\vec{R}$ & ${\tensor{X}}$ \\
\hline
reactive handed& $a$  & $\vec{O}$ & ${\tensor{Y}}$ \\
\hline
\end{tabular*}
\label{tab:3-tensor}
\caption{CEO in scalar-vector-tensor (SVT) classification, 
\emph{i.e.}, 3-tensor notation:
$4\times(1+3+5)=36$ observables.}
\end{table}
The fundamental space-time 4-tensor CEO can be written in terms of the 
three-dimensional
$\mathbf{E}$ and $\mathbf{B}$ vectors, \emph{i.e.},
3-tensors. This is convenient because it allows us to use intuitive physical  
quantities.
To systematize the 3D representation of the CEO, 
we will use a physical classification
where we organize the CEO into four groups,
which have been introduced briefly in the previous section:
the \emph{(active) total}, \emph{(active) handed}, 
\emph{reactive total}, and \emph{reactive handed} 
CEO parameter groups, respectively.  
In addition, we will use a coordinate-free 3D formalism and  
classify the CEO parameters according to rank, \emph{i.e.}, 
as
scalars, 3-vectors, and rank two 
3-tensors
(SVT classification). The 3D CEO are listed in Table \ref{tab:3-tensor}.
The CEO 3-tensors are defined as follows. 

The ``total'' parameters are:
\begin{align}
u=&T^{00}=  \left(|\mathbf{E}|^{2}+|\mathbf{B}|^{2}\right)/2\\
\vec{N}=&T^{i0}=  \Re\left\{ \CompConj{\mathbf{E}}\times\mathbf{B}\right\} \\
{\tensor{M}}=&T^{ij}= u\mathbf{1}_{3} -\Re\left\{ \CompConj{\mathbf{E}}\otimes\mathbf{E}+\CompConj{\mathbf{B}}\otimes\mathbf{B}\right\} 
\end{align}
where $\mathbf{1}_{3}$ is the identity matrix in three dimensions. 
This
is the 3D representation of the well-known energy-stress 4-tensor 
$T^{\alpha\beta}$, defined by Eq. (\ref{eq:T-4}).

The ``handed'' parameters are:
\begin{align}
h=&U^{00}=  \Im\left\{ \CompConj{\mathbf{E}}\cdot\mathbf{B}\right\} \\
\vec{S}=&U^{i0}=  -\frac{1}{2}\Im\left\{\left(\CompConj{\mathbf{E}}
\times\mathbf{E}+\CompConj{\mathbf{B}}\times\mathbf{B}\right)\right\}\\
{\tensor{C}}=&U^{ij}= h\mathbf{1}_{3} -\Im\left\{ \CompConj{\mathbf{E}}\otimes\mathbf{B}-\CompConj{\mathbf{B}}\otimes\mathbf{E}\right\} 
\end{align}
This
is the 3D representation of the handed energy-stress 4-tensor 
$U^{\alpha\beta}$, defined by Eq. (\ref{eq:U-4}).

The ``reactive total'' parameters are:
\begin{align}
l=&C_{+}=  \left(|\mathbf{E}|^{2}-|\mathbf{B}|^{2}\right)/2\\
\vec{R}=&Q^{i0}= -\Im\left\{ \CompConj{\mathbf{E}}\times\mathbf{B}\right\} \\
{\tensor{X}}=&W^{i0j0}= \frac{1}{2}\left(\Re\left\{ \CompConj{\mathbf{E}}\otimes\mathbf{E}
-\CompConj{\mathbf{B}}\otimes\mathbf{B}\right\}-\frac{2}{3}l\vec{1}_3\right) 
\end{align}
Contrary to the active, total and handed, parameter groups above, the reactive 
total parameter group have no single corresponding 4-tensor. 
Instead it is
composed of parts from three different CEO space-time tensors: 
the vacuum proper-Lagrangian defined by Eq. (\ref{eq:C+}), 
the reactive energy flux density from
Eq. (\ref{eq:Q-4}), and the generalized Stokes parameters corresponding
to the auto-correlated
$\vec{E}$ and $\vec{B}$ fields from Eq. (\ref{eq:W}).

The ``reactive handed''  parameters are:
\begin{align}
a=&C_{-}=  -\Re\left\{ \mathbf{\CompConj{E}}\cdot\mathbf{B}\right\}\\ 
\vec{O}=&\frac{1}{2}\epsilon^j_{kl}Q^{kl}=  
-\frac{1}{2}\Im\left\{\left(\CompConj{\mathbf{E}}
\times\mathbf{E}-\CompConj{\mathbf{B}}\times\mathbf{B}\right)\right\}\\
{\tensor{Y}}=&\frac{1}{2}\epsilon^j_{kl}W^{i0kl}= \frac{1}{2}\left(\Re\left\{ \CompConj{\mathbf{E}}
\otimes\mathbf{B}
+\CompConj{\mathbf{B}}\otimes\mathbf{E}\right\}-\frac{2}{3}a\vec{1}_3\right) 
\end{align}
Also for this parameter group, there is no single corresponding 4-tensor.
The reactive handed group is composed of parts from three CEO 
space-time tensors: the vacuum pseudo-Lagrangian, 
defined by Eq. (\ref{eq:C-}), the reactive handed energy flux density from
Eq. (\ref{eq:Q-4}), and the generalized Stokes parameters corresponding
to the cross-correlated
$\vec{E}$ and $\vec{B}$ fields from Eq. (\ref{eq:W}).

\section{CEO in two dimensions}
\label{sec:2D}
Up until now we have assumed that all three Cartesian components of
both the electric field, $\mathbf{E}$, and the magnetic 
field, $\mathbf{B}$, are measured. One may ask what happens if some
components are not measured; can all the 36 parameters of the CEO
be retained? Of course this is not possible, some information is certainly
lost in this case, but what one can do is to construct a set of parameters
analogous to CEO in two-dimensions.

Assume that we can measure the electric field and the magnetic 
field in a plane which we can say is the $xy$-plane without loss
of generality. Let the two-dimensional (2D) fields in this plane be denoted
$\mathbf{E}_{\mathrm{2D}}:=(E_{x},E_{y})$ and $\mathbf{B}_{\mathrm{2D}}:=(B_{x},B_{y})$,
and define the scalar product between 2D vectors as 
\begin{align}
\mathbf{E}_{\mathrm{2D}}\cdot\mathbf{B}_{\mathrm{2D}}=E_{x}B_{x}+E_{y}B_{y}
\end{align}

 and the cross product as 
 \begin{align}
\mathbf{E}_{\mathrm{2D}}\times\mathbf{B}_{\mathrm{2D}}=E_{x}B_{y}-E_{y}B_{x}
\end{align}

 and the direct product as 
\begin{align}
\mathbf{E}_{\mathrm{2D}}\otimes\mathbf{B}_{\mathrm{2D}}=\left(\begin{array}{cc}
E_{x}B_{x} & E_{x}B_{y}\\
E_{y}B_{x} & E_{y}B_{y}\end{array}\right)
\end{align}

We will however not need to consider all the components of the 2D
direct product since the 2-tensors we will consider are all symmetric
and traceless. Hence, we only want the parameters which correspond
to Pauli spin matrix components 
\begin{align}
\sigma_{x} & =\left(\begin{array}{cc}
0 & 1\\
1 & 0\end{array}\right)\\
\sigma_{z} & =\left(\begin{array}{cc}
1 & 0\\
0 & -1\end{array}\right)\end{align}
 The Pauli components can be extracted from a 2D matrix by matrix
multiplying by a Pauli spin matrix and then taking the trace, that
is
\begin{align}
\mathrm{Tr}\left\{ \left(\mathbf{E}_{\mathrm{2D}}\otimes\mathbf{B}_{\mathrm{2D}}\right)\sigma_{x}\right\}  & =\left(\mathbf{E}_{\mathrm{2D}}\otimes\mathbf{B}_{\mathrm{2D}}\right)\cdot\cdot\sigma_{x}\nonumber\\
 & =E_{x}B_{y}+E_{y}B_{x}\end{align}
where we have introduced the double scalar product, $\sigma_{x}\cdot\cdot\sigma_{y}$, see \cite{Lebedev03}.

We can derive a set of two-dimensional canonical electromagnetic parameters
from the full CEO by formally taking
\begin{align}
E_{z}\equiv B_{z}\equiv0
\end{align}
and discarding all the parameters that are identically zero. In this
way we obtain the following set, which we write in the coordinate-free
2D formalism introduced above. 

The ``total'' 2D parameters are: 
\begin{align}
u_{2D} & =\left(\left|\mathbf{E}_{\mathrm{2D}}\right|^{2}+\left|\mathbf{B}_{\mathrm{2D}}\right|^{2}\right)/2\nonumber\\
 & =\left(\left|E_{x}\right|^{2}+\left|E_{y}\right|^{2}+\left|B_{x}\right|^{2}+\left|B_{y}\right|^{2}\right)/2\\
N_{z} & =\Re\left\{ \mathbf{E}_{\mathrm{2D}}\times\CompConj{\mathbf{B}}_{\mathrm{2D}}\right\} \nonumber\\
 & =\Re\left\{ E_{x}\CompConj{B}_{y}-E_{y}\CompConj{B}_{x}\right\} \\
M_{\sigma_{z}} & =\Re\left\{ \mathbf{E}_{\mathrm{2D}}\otimes\CompConj{\mathbf{E}}_{\mathrm{2D}}+\mathbf{B}_{\mathrm{2D}}\otimes\CompConj{\mathbf{B}}_{\mathrm{2D}}\right\} \cdot\cdot\sigma_{z}/2\nonumber\\
 & =\left(\left|E_{x}\right|^{2}-\left|E_{y}\right|^{2}+\left|B_{x}\right|^{2}-\left|B_{y}\right|^{2}\right)/2\\
M_{\sigma_{x}} & =\Re\left\{ \mathbf{E}_{\mathrm{2D}}\otimes\CompConj{\mathbf{E}}_{\mathrm{2D}}+\mathbf{B}_{\mathrm{2D}}\otimes\CompConj{\mathbf{B}}_{\mathrm{2D}}\right\} \cdot\cdot\sigma_{x}/2\nonumber\\
 & =\Re\left\{ E_{x}\CompConj{E}_{y}+B_{x}\CompConj{B}_{y}\right\} \end{align}

The ``handed'' 2D parameters are:
\begin{align}
h_{2D} & =\Im\left\{ \mathbf{E}_{\mathrm{2D}}\cdot\CompConj{\mathbf{B}}_{\mathrm{2D}}
\right\}\nonumber \\
 & =\Im\left\{ E_{x}\CompConj{B}_{x}+E_{y}\CompConj{B}_{y}\right\} \\
S_{z} & =\Im\left\{ \mathbf{E}_{\mathrm{2D}}\times\CompConj{\mathbf{E}}_{\mathrm{2D}}
+\mathbf{B}_{\mathrm{2D}}\times\CompConj{\mathbf{B}}_{\mathrm{2D}}
\right\}\nonumber \\
 & =\Im\left\{ E_{x}\CompConj{E}_{y}+B_{x}\CompConj{B}_{y}\right\} \\
C_{\sigma_{z}} & =\Im\left\{ \mathbf{E}_{\mathrm{2D}}\otimes\CompConj{\mathbf{B}}_{\mathrm{2D}}-\mathbf{B}_{\mathrm{2D}}\otimes\CompConj{\mathbf{E}}_{\mathrm{2D}}
\right\} \cdot\cdot\sigma_{z}/2\nonumber\\
 & =\Im\left\{ E_{x}\CompConj{B}_{x}-E_{y}\CompConj{B}_{y}\right\} \\
C_{\sigma_{x}} & =\Im\left\{ \mathbf{E}_{\mathrm{2D}}\otimes\CompConj{\mathbf{B}}_{\mathrm{2D}}-\mathbf{B}_{\mathrm{2D}}\otimes\CompConj{\mathbf{E}}_{\mathrm{2D}}\right\} 
\cdot\cdot\sigma_{x}/2\nonumber\\
 & =\Im\left\{ E_{x}\CompConj{B}_{y}+E_{y}\CompConj{B}_{x}\right\} \end{align}

The ``reactive total'' 2D parameters are:
\begin{align}
l_{2D} & =\left(\left|\mathbf{E}_{\mathrm{2D}}\right|^{2}
-\left|\mathbf{B}_{\mathrm{2D}}\right|^{2}\right)/2\nonumber\\
 & =\left(\left|E_{x}\right|^{2}+\left|E_{y}\right|^{2}-\left|B_{x}\right|^{2}-\left|B_{y}\right|^{2}\right)/2\\
R_{z} & =\Im\left\{ \mathbf{E}_{\mathrm{2D}}\times\CompConj{\mathbf{B}}_{\mathrm{2D}}
\right\}\nonumber \\
 & =\Im\left\{ E_{x}\CompConj{B}_{y}-E_{y}\CompConj{B}_{x}\right\} \\
X_{\sigma_{z}} & =\Re\left\{ \mathbf{E}_{\mathrm{2D}}\otimes\CompConj{\mathbf{E}}_{\mathrm{2D}}-\mathbf{B}_{\mathrm{2D}}\otimes\CompConj{\mathbf{B}}_{\mathrm{2D}}\right\} 
\cdot\cdot\sigma_{z}/2\nonumber\\
 & =\left(\left|E_{x}\right|^{2}-\left|E_{y}\right|^{2}-\left|B_{x}\right|^{2}+\left|B_{y}\right|^{2}\right)/2\\
X_{\sigma_{x}} & =\Re\left\{ \mathbf{E}_{\mathrm{2D}}\otimes\CompConj{\mathbf{E}}_{\mathrm{2D}}-\mathbf{B}_{\mathrm{2D}}\otimes\CompConj{\mathbf{B}}_{\mathrm{2D}}\right\} 
\cdot\cdot\sigma_{x}/2\nonumber\\
 & =\Re\left\{ E_{x}\CompConj{E}_{y}-B_{x}\CompConj{B}_{y}\right\} \end{align}

The ``reactive handed'' 2D parameters are:
\begin{align}
a_{2D} & =\Re\left\{ \mathbf{E}_{\mathrm{2D}}\cdot\CompConj{\mathbf{B}}_{\mathrm{2D}}
\right\}\nonumber \\
 & =\Re\left\{ E_{x}\CompConj{B}_{x}+E_{y}\CompConj{B}_{y}\right\} \\
O_{z} & =\Im\left\{ \mathbf{E}_{\mathrm{2D}}\times\CompConj{\mathbf{E}}_{\mathrm{2D}}
-\mathbf{B}_{\mathrm{2D}}\times\CompConj{\mathbf{B}}_{\mathrm{2D}}\right\}\nonumber \\
 & =\Im\left\{ E_{x}\CompConj{E}_{y}-B_{x}\CompConj{B}_{y}\right\} \\
Y_{\sigma_{z}} & =\Re\left\{ \mathbf{E}_{\mathrm{2D}}\otimes\CompConj{\mathbf{B}}_{\mathrm{2D}}+\mathbf{B}_{\mathrm{2D}}\otimes\CompConj{\mathbf{E}}_{\mathrm{2D}}\right\} 
\cdot\cdot\sigma_{z}/2\nonumber\\
 & =\Re\left\{ E_{x}\CompConj{B}_{x}-E_{y}\CompConj{B}_{y}\right\} \\
Y{}_{\sigma_{x}} & =\Re\left\{ \mathbf{E}_{\mathrm{2D}}\otimes\CompConj{\mathbf{B}}_{\mathrm{2D}}+\mathbf{B}_{\mathrm{2D}}\otimes\CompConj{\mathbf{E}}_{\mathrm{2D}}\right\} 
\cdot\cdot\sigma_{x}/2\nonumber\\
 & =\Re\left\{ E_{x}\CompConj{B}_{y}+E_{y}\CompConj{B}_{x}\right\} \end{align}

We can associate names with these parameters as listed in Table \ref{CEO2Dnames}.
The first four parameters, which we call the {}``total'' 2D CEO
parameters are all well known. 
These parameters are also known by different
names, e.g., the total energy flux is also known as the Poynting vector
($z$-component), and the total energy stress is known as the Maxwell
stress tensor (difference of diagonal components and off-diagonal
component). The remaining three sets of 2D CEO parameters are less well known.
We will not be able to provide a full physical interpretation of each
of these parameters; indeed their role in space plasma physics is
yet to be fully explored. We will only mention that the {}``handed''
parameters involve spin (helicity, chirality, polarization) 
weighted energy, i.e., the energy
of the right-hand wave modes are weighted positively and the energy
of left-hand wave modes are weighted negatively, and these weighted
energies are then added. Its flux corresponds to the concept of ellipticity
and for the case of vacuum, it is numerically equivalent to Stokes
$V$ parameter. The reactive energy densities come in two groups: 
the ``reactive total''
and the ``reactive handed''  2D CEO parameter groups. From the ``reactive total'' group,
we now recognize the reactive energy flux density, as well as the EM Stokes 
$Q$ and $U$ parameters,
which here are of the auto-type; the vacuum proper-Lagrangian needs no further 
introduction. The ``reactive handed'' group contain the handed counterparts of the
reactive energy flux density and EM Stokes parameters, which here are
of the cross-typer; the vacuum pseudo-Lagrangian is well-known.

\begin{table}[t]
\begin{tabular*}{\columnwidth}{@{\extracolsep{\fill}}|cl|}
\hline 
Symbol&
Detailed Name\tabularnewline
\hline 
\hline 
$u_{2D}$&
Total energy\tabularnewline
\hline 
$N_{z}$$ $&
Total energy flux\tabularnewline
\hline 
$M_{\sigma_{z}}$&
Total energy stress $\sigma_{z}$component\tabularnewline
\hline 
$M_{\sigma_{x}}$&
Total energy stress $\sigma_{x}$component\tabularnewline
\hline
\hline 
$h_{2D}$&
Handed energy\tabularnewline
\hline 
$S_{z}$&
Handed energy flux\tabularnewline
\hline 
$C_{\sigma_{z}}$&
Handed energy stress $\sigma_{z}$-component\tabularnewline
\hline 
$C_{\sigma_{x}}$&
Handed energy stress $\sigma_{x}$-component\tabularnewline
\hline
\hline 
$l_{2D}$&
Vacuum proper-Lagrangian\tabularnewline
\hline 
$R_{z}$&
Reactive energy flux\tabularnewline
\hline 
$X_{\sigma_{z}}$&
EM Stokes parameter Q auto-type\tabularnewline
\hline 
$X_{\sigma_{x}}$&
EM Stokes parameter U auto-type\tabularnewline
\hline
\hline 
$a_{2D}$&
Vacuum pseudo-Lagrangian\tabularnewline
\hline 
$O_{z}$&
Reactive handed energy flux\tabularnewline
\hline 
$Y_{\sigma_{z}}$&
EM Stokes parameter Q cross-type\tabularnewline
\hline 
$Y{}_{\sigma_{x}}$&
EM Stokes parameter U cross-type\tabularnewline
\hline
\end{tabular*}
\caption{\label{CEO2Dnames}Naming scheme for the 2D electromagnetic second-order
parameters. All parameters are implicitly densities and two-dimensional.}
\end{table}
\begin{figure*}
\includegraphics[width=1\textwidth,keepaspectratio]{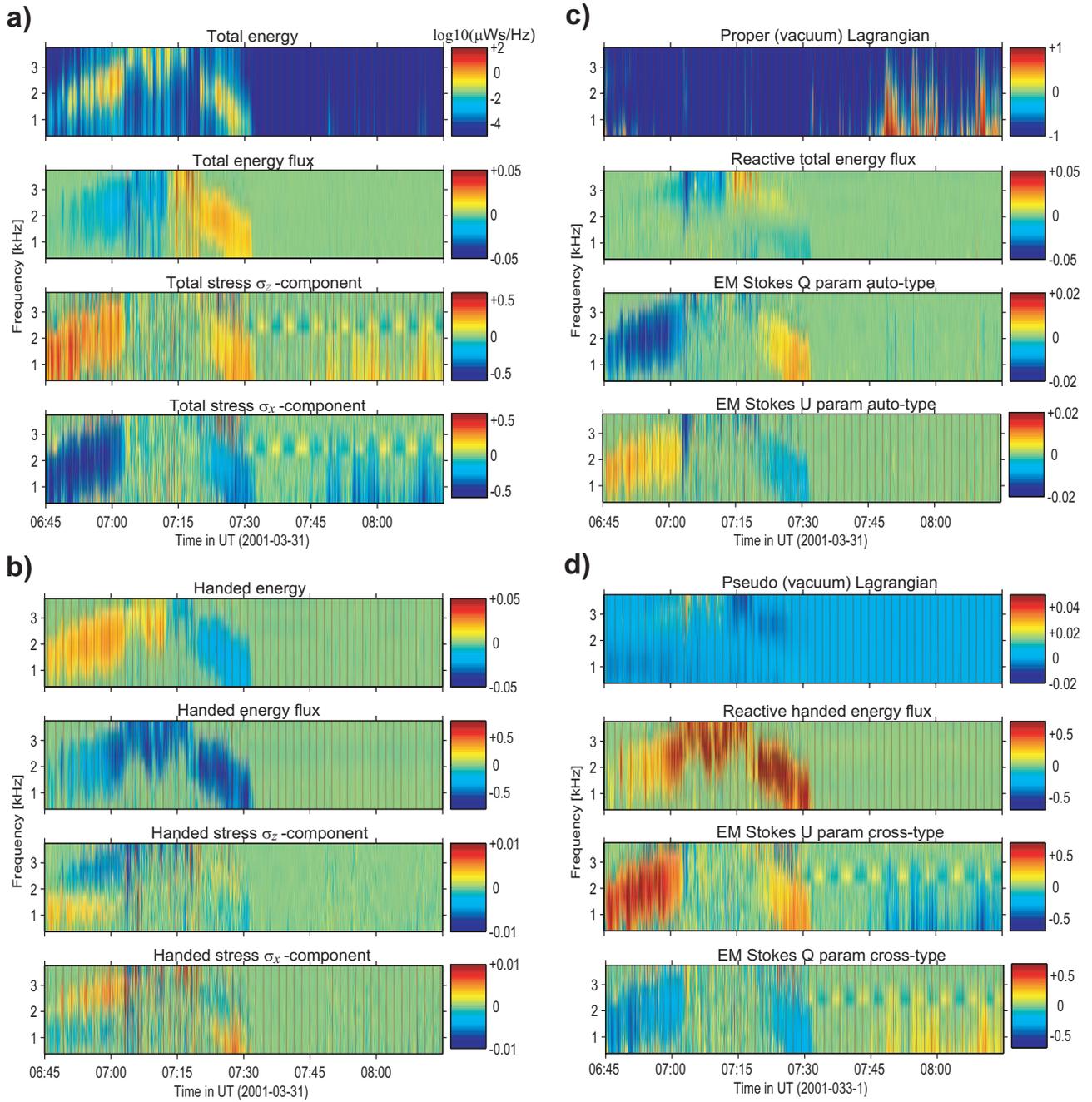}

\caption{\label{fig:Parrot03B3_4x4}Example dynamic spectra of the 16 normalized
two-dimensional CEO parameters. The parameters were computed from
STAFF-SA data from Cluster space-craft 2 using the ISDAT database
system. The following normalization has been applied: each parameter
has been divided by the total energy except the total energy itself.
Thus all spectral values are in dimensionless unit except for the
total energy. This Figure can be compared with Fig. 1 in \cite{Parrot03a}.
The 16 parameters are subdivided into a) the total energy parameters,
b) the handed energy parameters, c) the reactive total energy parameters,
and d) the reactive handed energy parameters.
Note that all the quantities are purely electromagnetic
in origin and so do not refer to contributions from the plasma.}
\label{fig:2D-CEO}
\end{figure*}

\section{Application of CEO to Cluster data}
Let us demonstrate that the CEO parameters can easily be computed
from actual data. Assuming that we have measurements from a vector
magnetometer and an electric field instrument, all that is required
is to auto/cross-correlate all measured components and then form the
appropriate linear combination introduced above.

As an example we will consider the STAFF-SA dataset on the Cluster-II
space-craft mission \cite{Escoubet97}. 
The STAFF-SA instrument \cite{Cornilleau-Wehrlin1997} is well suited for
the CEO parameters since it outputs auto/cross-correlation of electric
and magnetic field components; however as Cluster does not measure
one of the electric field components (namely the component normal
to the spin-plane of the space-craft) we can only use the 2D
version of the CEO introduced in the previous section. 

For this particular example, we re-process the high-band part of STAFF-SA data from
an event discussed in \cite{Parrot03a} from 2001-03-31 UT. In Fig.~1 of this
paper, Parrot \emph{et al} display certain parameters based on the STAFF-SA
data computed using a numerical software package called PRASSADCO;
see \cite{Santolik2003}. The interesting feature of the 2D CEO parameters is
that they are the complete set of electromagnetic field observables
in the spin-plane of the space-craft; and indeed, they use up all the
parameters in the STAFF-SA dataset expect for the magnetic field in
the spin direction. Each CEO is a distinct physical quantity and examination
of the panels in Fig. \ref{fig:Parrot03B3_4x4} indicates that this is indeed the case,
since besides showing a 
common chorus feature (the arch to the left in each panel) there are unique
points in each of the panels.

Besides being a complete description of the electromagnetic observables,
the fact that the CEO parameters are based on parameters that conform
with the physics of space-time means that we can expect physical phenomenon
to be measured properly. Seeing as how the CEO parameters have not
been explicitly measured in the past, we can expect that their future
use may lead to new physical insights, especially since several of the parameters
are completely new to space-physics. As an example consider again
the data shown in Fig \ref{fig:Parrot03B3_4x4}. It is interesting
to note that the reactive total energy flux is only significant close
to the equator; this implies that the equator is the source region
for the chorus events, since reactive energy flux is typically large
close to radiating objects due to large standing energy fields. One
can also see a modulation at $2.5$ kHz in the EM Stokes parameters.
If this is a physical phenomenon it would be indicative of Faraday
rotation. Also there seems to be frequency dispersion in the handed
stress since its components changes sign with frequency. Finally, the
handed energy clearly shows the handedness of the chorus
emissions on its own, without recourse to the sign of the total energy
flux. 

\section{Conclusion}
The proposed CEO parameters conveniently organize the measurements of the full EM field.
Furthermore, they are physically meaningful quantities, \emph{i.e.} they
\begin{itemize}
\item have conservation laws
\item transform as geometric (Minkowski space-time) objects
\item are mathematically unique (they are irreducible tensors)
\item retain all information, \emph{i.e.} nothing is lost 
(linear transformation back to full sixtor form exists)
\item enables considerable data reduction (through parameter subset selection)
\item have clear despinning properties (\emph{e.g.} scalar quantities do not need despinning!)
\item are all real valued
\item provides useful decomposition of the 36 second order EM components into twelve
3-tensor quantities
\item reveals some new physical parameters describing EM waves: opening for 
new physical insights.
\end{itemize}

\section*{Acknowledgments}
We would like to thank the participants and the organization of the 
Solar Orbiter Workshop II for their valuable input to this work. 
Many of the ideas developed in this paper were sprung from
presentations and discussions
during the workshop.
Specifically, we would like to thank Professor Xenophon Moussas 
from the University of Athens, for his great hospitality and
support of our work.
We would also like to thank Dr.
Ond\v{r}ej Santol\'{i}k, from
Charles University in Prague, and  
Mr. Christopher Carr, from
Imperial College in London, 
for their
valuable comments and suggestions during the poster session.


\end{document}